\documentclass[twocolumn,showpacs,preprintnumbers,amsmath,amssymb]{revtex4}
\usepackage{graphicx}% Include figure files
\usepackage{dcolumn}% Align table columns on decimal point
\usepackage{bm}% bold math

\begin{document}
\title{Giant anisotropic photocurrent modulated by strain in type-II Weyl semimetal T$_d$-MoTe$_2$}
\author{Xinru Wang$^{1}$}
\author{Ying Ding$^{1}$}
\author{M. N. Chen$^{1}$}
\author{Z. B. Siu$^{2}$}
\author{Mansoor B. A. Jalil$^{2}$}
\email{elembaj@nus.edu.sg}
\author{Yuan Li$^{1}$}
\email{liyuan@hdu.edu.cn}
\affiliation{$^1$
Department of Physics, Hangzhou Dianzi University, Hangzhou, Zhejiang 310018, China}
\affiliation{$^2$ Computational Nanoelectronics and Nano-device Laboratory, Electrical and Computer
Engineering Department, National University of Singapore, 4 Engineering Drive 3,
Singapore 117576, Singapore}

\begin{abstract}
We build a Cu-MoTe$_2$-Cu device model and use first-principles density functional theory to study the transport properties of single-layer T$_d$-MoTe$_2$. We obtained the effect of strain on the energy band structure, transport properties, and photocurrent.
The strain-induced photocurrent shows an anisotropy that reflects the modulation of the energy bands, including the Weyl point, by strain. The photocurrent can be suppressed to almost zero when the strain is applied along the vacuum direction. In contrast, the photocurrent can be significantly increased when the strain is applied along the transport direction. The transport properties and magnitude of the photocurrent in the MoTe$_2$-based device can be effectively modulated by adjusting the strength and direction of the strain.
\end{abstract}
\date{\today}
\maketitle

\section{Introduction}
The continuous in-depth research on the theory and fabrication technology of two-dimensional material devices in recent years has witnessed a development from the initial silicon materials to graphene-based layered nanomaterials~\cite{Rao,Mondal}, and subsequently to the current transition metal chalcogenide materials. Two-dimensional transition-metal dichalcogenides (TMDs)~\cite{Wilson} have quickly become a research hotspot in the field of materials because of their high carrier mobility, appropriate band gap, large switching ratio, and layer-dependent band gap~\cite{Xin,Luxa,Zeng}. MoTe$_2$ is an important part of the TMD family. MoTe$_2$ exists in three phases~\cite{Dawson}, namely, the hexagonal (2H, semiconductor), monoclinic (1T', metal), and octahedral (T$_d$, type-II Weyl semimetal) phases. Our main research subject is T$_d$-MoTe$_2$, which has been theorized to be an example of a Type-II Weyl semimetal~\cite{Guguchia,Jiang,Deng}.

The Type-II Weyl semimetal is a generalization of the Weyl semimetal. Its defining characteristic is the inclination of the Dirac cone near the Weyl point, which results in a corresponding electronic dispersion relation that does not satisfy Lorentz invariance near the Weyl point~\cite{Soluyanov}. T$_d$-MoTe$_2$ has rich physical properties which include superconductivity~\cite{Qi}, extremely large magnetoresistance (XMR)~\cite{Dong}, and topological semi-metal properties. Because single-layer T$_d$-MoTe$_2$ is predicted to host the quantum spin Hall (QSH) insulating state, this material has attracted much attention in the study of condensed matter physics.

In this study, we calculate the transport properties of two-dimensional MoTe$_2$-based devices based on quantum transport simulations in which density functional theory (DFT) is combined with the non-equilibrium Green's function (NEGF)~\cite{Jauho,Wang,Wang2} formalism. We analyze the effect of the strain on the  energy band diagram, transmission spectrum, and photocurrent of layered T$_d$-MoTe$_2$ to provide a theoretical basis for subsequent experimental research.

\begin{figure}[b]
\centering
\includegraphics[scale= 0.4]{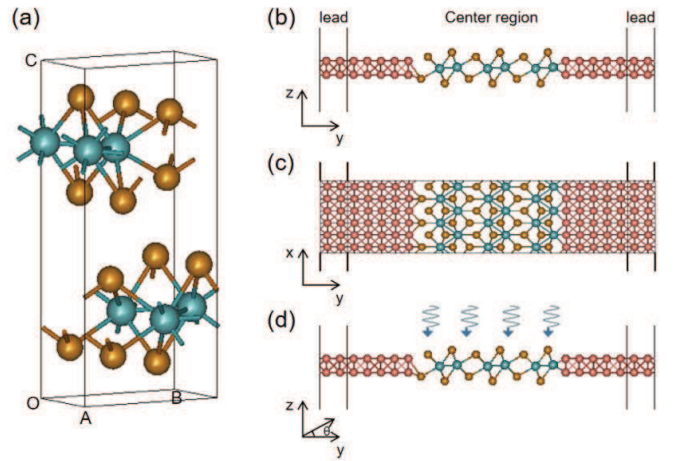}
\caption{\label{fig:cell} (a) The unit cell of Td-MoTe$_2$. (b) Side and (c) top view of Cu-MoTe$_2$-Cu transport model. (d) The Td-MoTe$_2$ in the central region is irradiated by linearly polarized light.}
\end{figure}

\section{Theoretical model and method}
We model a device with the Cu-MoTe$_2$-Cu configuration in which a Td-MoTe$_2$  central region is sandwiched between two copper leads. Fig.~\ref{fig:cell}(a) shows the unit cell of T$_d$-MoTe$_2$. The space group of T$_d$-MoTe$_2$ is Pmn2$_{1}$(No.31) and the lattice parameters are $a=3.477{\AA}$, $b=6.335{\AA}$, and $c=13.889{\AA}$. These datum are consistent with previously reported structure ~\cite{Qi}. The device model is shown in Fig.~\ref{fig:cell}(b) and (c). The $y$ direction is the transport direction, and the $x$ direction is the transverse periodic direction which extends to $\pm\infty$ so that the device lies on the $x-y$ plane. The $z$ axis is the vacuum direction. A 40{\AA} vacuum region is used to separate the devices in this direction. As shown in Fig.~\ref{fig:cell}(d), strain and linearly polarized light are applied to the T$_d$-MoTe$_2$ in the central region.
\begin{figure*}[t]
\centering
\includegraphics[scale=0.35]{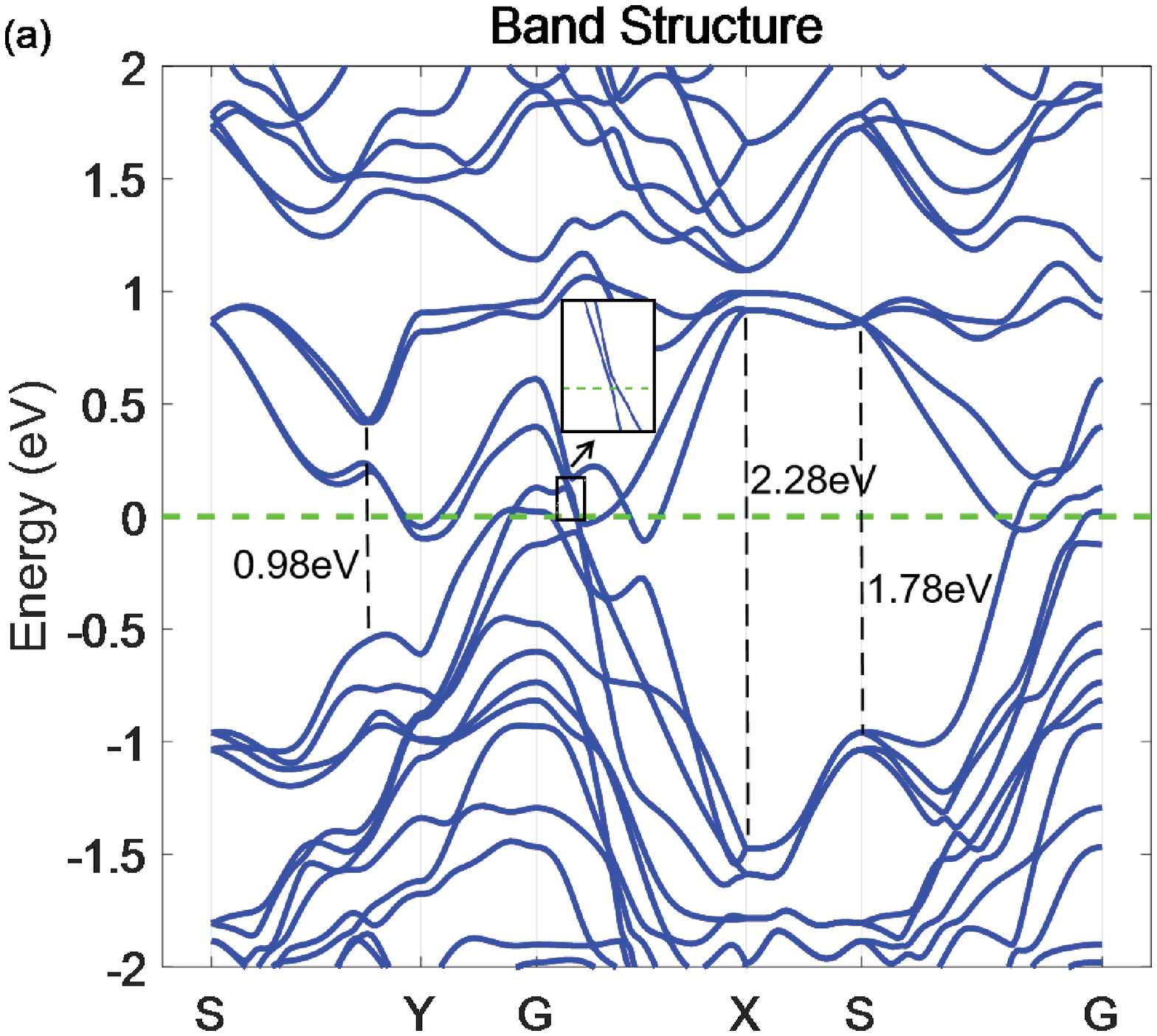}%
	\includegraphics[scale=0.35]{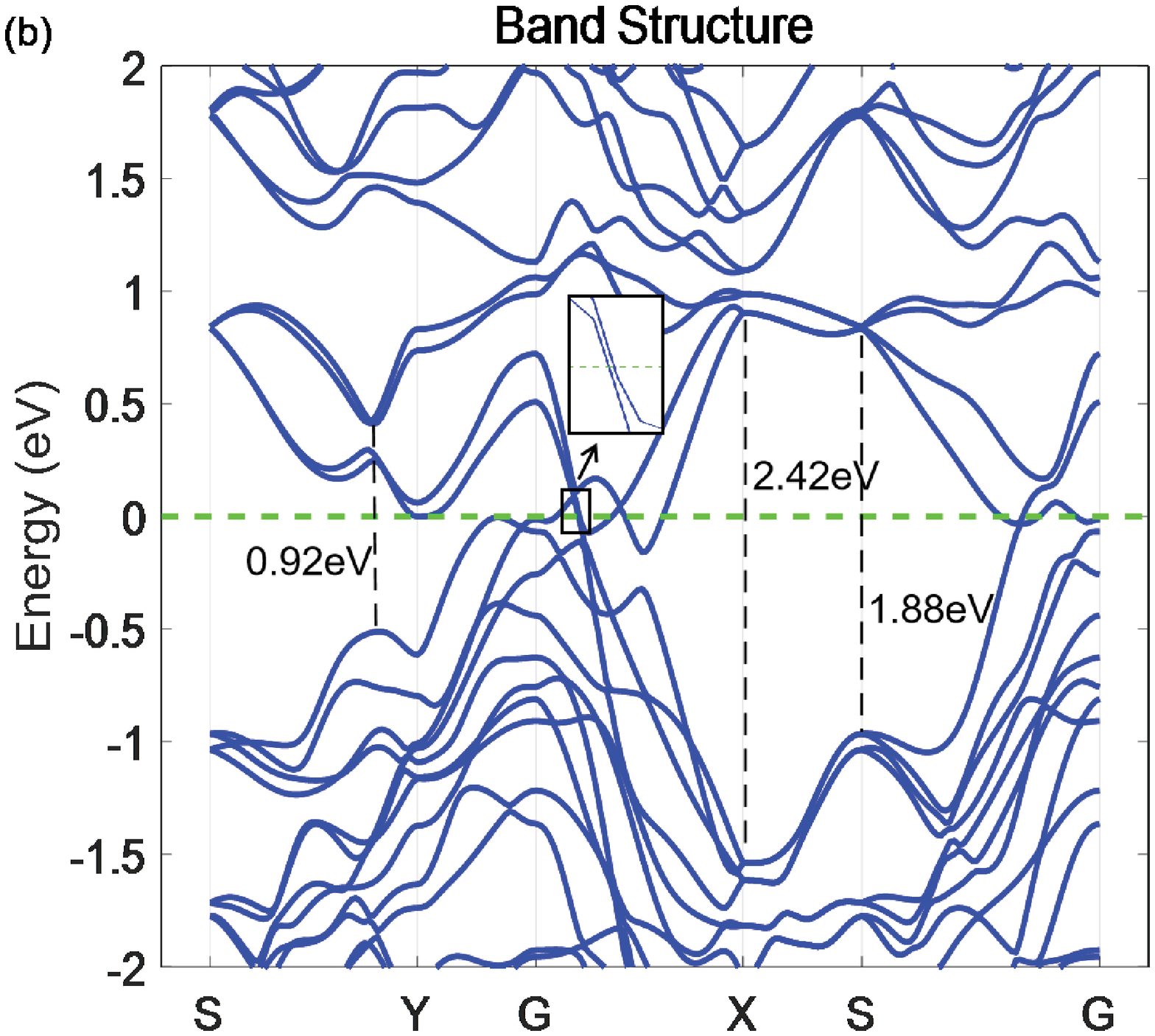}%
	
	\includegraphics[scale=0.35]{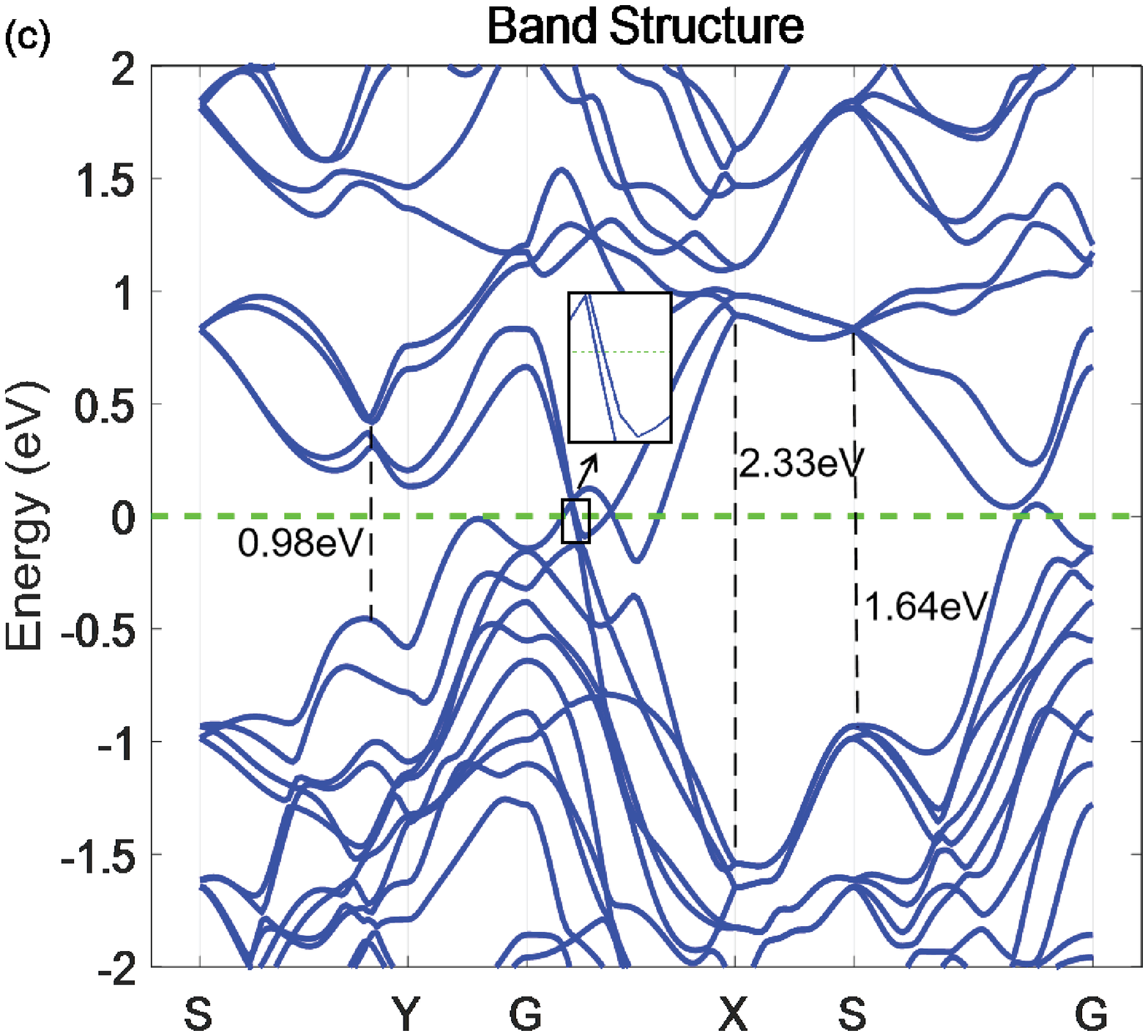}%
	\includegraphics[scale=0.35]{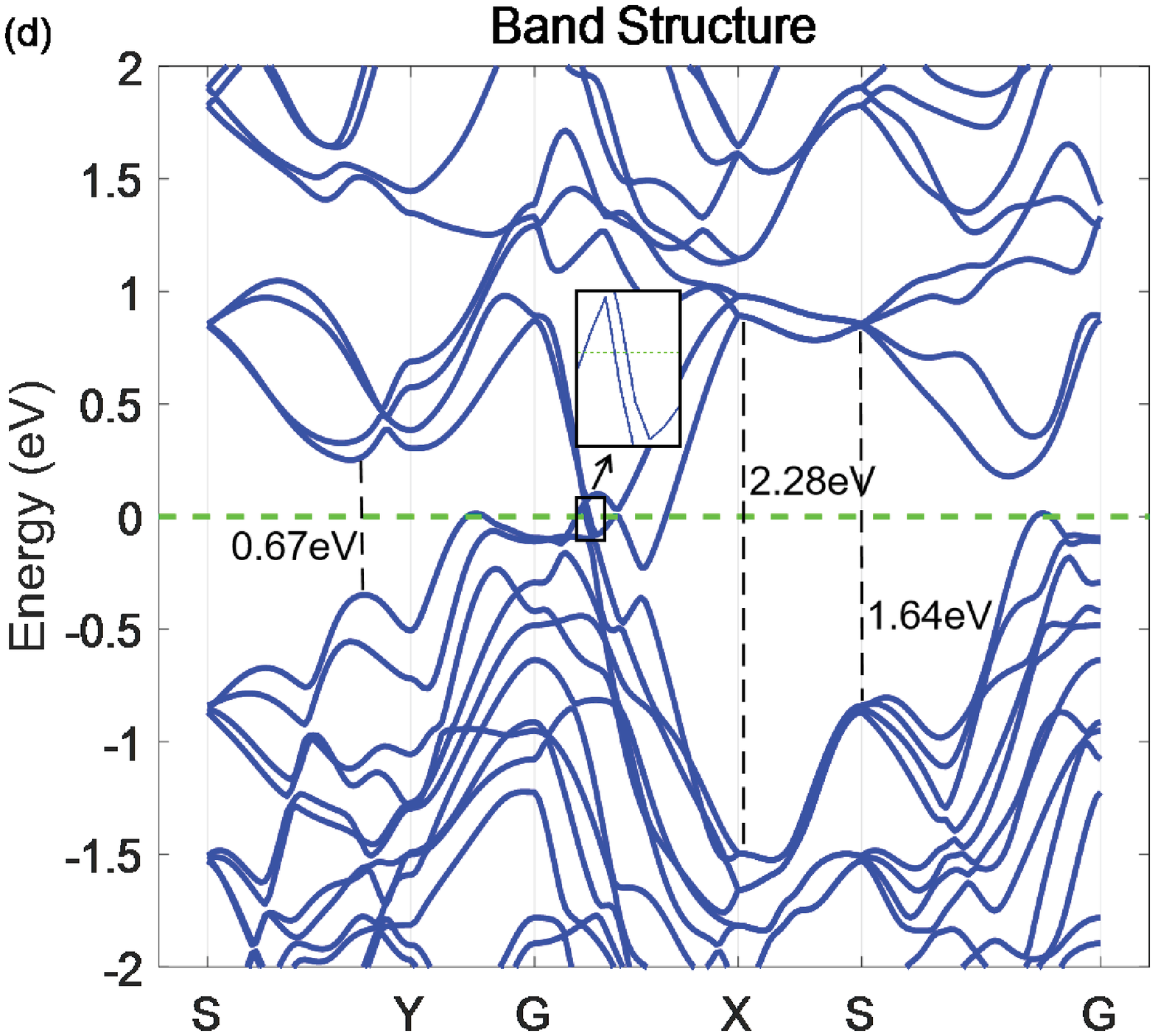}%
\caption{\label{fig:0Band} Energy band structure of Td-MoTe$_2$ for different strain strengths exerted along the $x$ direction: (a) $\varepsilon=0$, (b) $\varepsilon=3\%$, (c) $\varepsilon=6\%$, and (d) $\varepsilon=9\%$.}
\end{figure*}

The DFT-based software package Nanodcal ~\cite{Taylor} was used for transport calculations in this study. In the quantum transport calculations, the charge density was obtained through self-consistent calculations and the conductivity in turn calculated from the charge density. The LDA$_{-}$PZ81 generalized gradient approximation was used in the DFT calculations to describe the exchange-correlation energy. The plane wave basis set cutoff energy was set as 100 Hartrees (1 Hartree = 27.21 eV). The Brillouin zone of T$_d$-MoTe$_2$ was sampled with a $10\times11\times1$ k-mesh for geometry optimization and the self-consistent calculations. The self-consistent threshold for the total energy of the system was $10^{-4}$ eV and the strain of the system converged to 0.04 $eV/A$ in the relaxation calculation.

The strain is applied to the T$_d$-MoTe$_2$ material in the central region. The wavy lines in Fig.~\ref{fig:cell}(d) schematically represent the illumination of linearly polarized light. The polarized light is incident on the $x-y$ plane  and polarized at an angle $\theta$ with respect to the transport direction. When the T$_d$-MoTe$_2$ central region is irradiated by linearly polarized light, a photocurrent is generated.  The photocurrent can be written as~\cite{Xie,Moayed,Luo}
\begin{widetext}
\begin{eqnarray}
J_{L}^{\left ( ph\right )}&=&\frac{ie}{h}\int  \left \{
\cos^{2}\theta \mathrm{Tr}\left \{\Gamma_{L}\left[G_{1}^{<\left( ph\right )}+f_{L}\left( G_{1}^{>\left( ph\right )}-G_{1}^{<\left( ph\right )}\right )\right ]\right \} \right.+\sin^{2}\theta \mathrm{Tr}\left \{\Gamma _{L}\left [ G_{2}^{<\left ( ph\right )}+f_{L}\left ( G_{2}^{>\left ( ph\right )}-G_{2}^{<\left ( ph\right )}\right )\right ]\right \}\nonumber\\
&+&2 \sin\left ( 2\theta \right ) \mathrm{Tr}\left \{\Gamma _{L}\left [ G_{3}^{<\left ( ph\right )}+f_{L}\left ( G_{3}^{>\left ( ph\right )}-G_{3}^{<\left ( ph\right )}\right )\right ]\right \} \Big \}\ \mathrm{d}E, \nonumber\\
\end{eqnarray}
\end{widetext}
where
\begin{eqnarray}
 G_1^{ > ( < )ph} &=& \sum\limits_{\alpha ,\beta  = x,y,z} {{C_0}NG_0^r} {e_{1\alpha }}p_\alpha ^ \dag G_0^{ > ( < )}{e_{1\beta }}{p_\beta }G_0^a, \nonumber\\
 G_2^{ > ( < )ph} &=& \sum\limits_{\alpha ,\beta  = x,y,z} {{C_0}NG_0^r} {e_{2\alpha }}p_\alpha ^ \dag G_0^{ > ( < )}{e_{2\beta }}{p_\beta }G_0^a, \nonumber\\
 G_3^{ > ( < )ph} &=& \sum\limits_{\alpha ,\beta  = x,y,z} { {{C_0}} NG_0^r({e_{1\alpha }}p_\alpha ^ \dag G_0^{ > ( < )}{e_{2\beta }}{p_\beta }}\nonumber\\
 &+&{{e_{2\alpha }}p_\alpha ^ \dag G_0^{ > ( < )}{e_{1\beta }}{p_\beta })G_0^a}.
\end{eqnarray}
\begin{figure*}[t]
\centering
\includegraphics[scale=0.5]{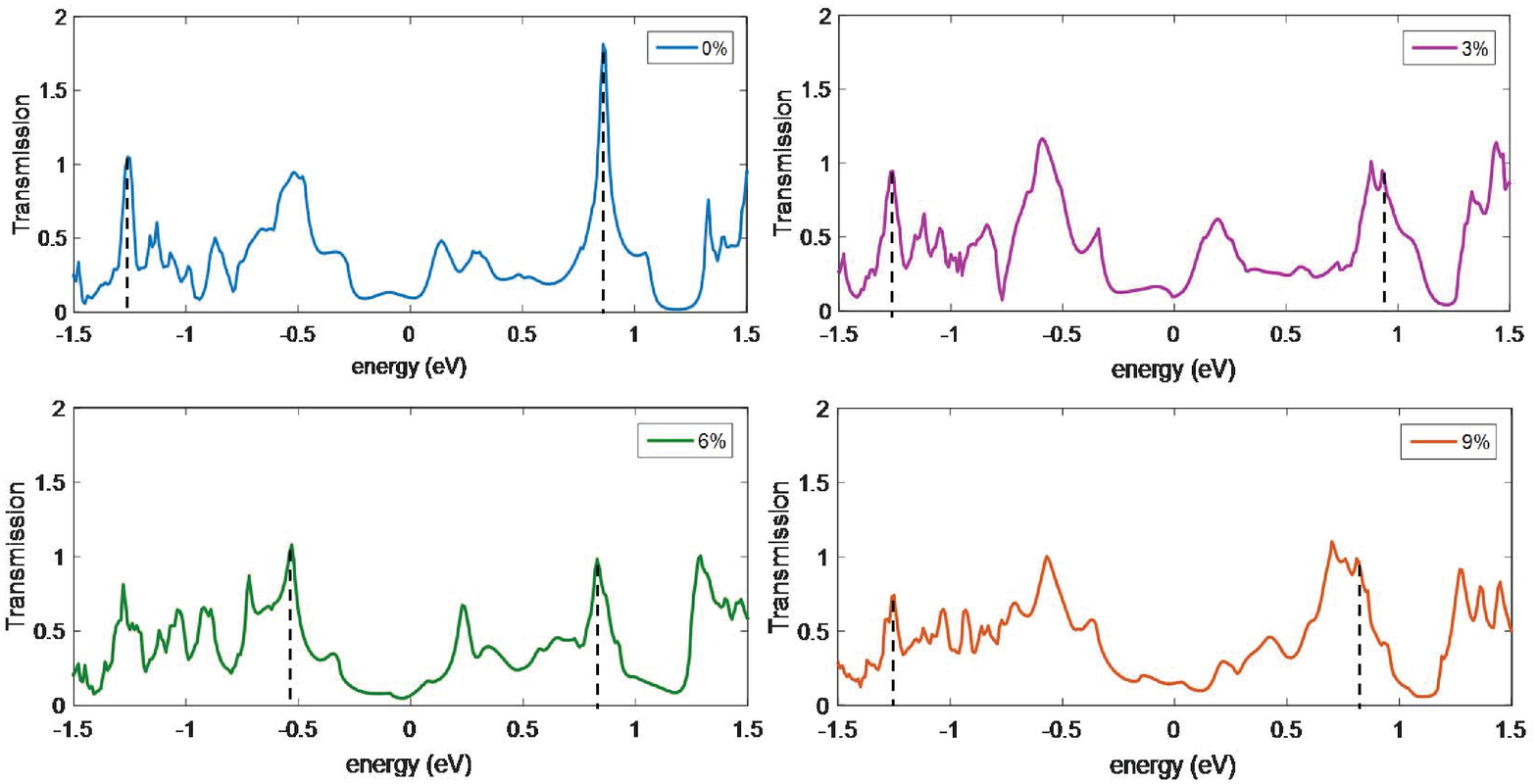}
\caption{\label{zttrans} The electron transmission plotted as a function of the Fermi energy for different amplitudes of the strain applied along $y$ axis.}
\end{figure*}

Here $G_{1,2,3}^{ > ( < )ph}$ is the greater or lesser Green's function of the electron-photon interaction, which is dependent on the photon frequency $\omega$, and ${f_L}$ is the Fermi-Dirac distribution function.  ${C_0} = \frac{{{e^2}\hbar \sqrt {{\mu _r}{\varepsilon _r}} }}{{2N{m^2}\omega \varepsilon c}}{I_\omega }$ where $m$ is the bare electron mass, the photon flux ${I_\omega }$ is the number of photons per unit area per unit time, $N$ is the number of photons, $\varepsilon$ is the dielectric constant, and ${\varepsilon _r}$ is the  relative dielectric constant. In our calculations, the polarization vector $\hat e = \cos \theta {\hat e_1} + \sin \theta {\hat e_2}$ describes the polarization of the incident light. The unit vectors $\hat{e}_1$ and $\hat{e}_2$ are set along the $y$ and $x$ directions, respectively. The photocurrent can be normalized as $J = J_L^{(ph)}/e{I_\omega }$.

All the calculations were performed using the quantum transport package in Nanodcal, which is based on DFT. The atomic cores were defined using non-local pseudopotentials and the interactions between the T$_d$-MoTe$_2$ centre region and the copper leads of the device calculated using NEGF. By combining DFT and NEGF, the calculation time for NEGF is shortened, and the accuracy of the DFT calculation results is improved.
The obtained band structure of MoTe$_2$ along several high-symmetry lines in the Brillouin zone are shown in Fig.~\ref{fig:0Band}. According to the results of Soluyanov et al.~\cite{Soluyanov}, the Type-II Weyl node results in a non-closed contact between the conduction band (CB) and valence band (VB) and an open Fermi surface. As shown in Fig.~\ref{fig:0Band}(a) (see inset), there is a Weyl point between the high-symmetry points G and X, which is consistent with the characteristics of the Type-II Weyl semimetal. In addition, a band crossing region can be observed along the path S-G. There exist three distinct values of energy separation, viz. $0.98$, $2.28$, and $1.78$ eV, between the local energy extrema of the CB and VB. The following calculation results show that strain can break the energy crossings at the Weyl points in T$_d$-MoTe$_2$.

\section{Results and discussion}
The strain is first applied to T$_d$-MoTe$_2$ along the $y$ direction (transport direction). The strength of the strain is expressed as $\zeta=(b-b_{0})/b_{0}$ where $b_{0}$ and $b$ are the lattice constants of the model in the absence and presence of the strain, respectively. The maximum value of the strain used in the calculations was $\zeta=10\%$ to maintain a realistic structure for T$_d$-MoTe$_2$. Figs.~\ref{fig:0Band}(b)-(d) show the energy bands obtained when 3$\%$, 6$\%$ and 9$\%$ strain are applied along the transport direction. At 3$\%$ strain, the bands that form the Weyl point begin to separate because of the strain although the energy bands have no distinct change. The energy level spacings are correspondingly changed. For example, the left level spacing decreases from $0.98$ eV to $0.92$ eV. When the strain increases to 6$\%$, the bands forming the Weyl point further separate and the crossing of the energy bands along the path S-G is also broken by the effect of the strain. With the strain increases to 9$\%$, the Weyl point is completely broken and a large energy separation appears between the original crossing energy bands. The left and right energy level spacings are significantly decreased to about $0.67$ eV and $1.64$ eV, respectively.

To investigate the effect of strain on the transport property, we study the transmission of electrons in the Cu-MoTe$_2$-Cu device under the action of the strain applied along the $y$ direction. Fig.~\ref{zttrans} shows the variation of the transmission with the strength of the strain. When $\zeta=0$, there exist three transmission peaks near $E=-1.3$, $-0.5$, and $0.9$ eV. This implies that there are more quantum states that contribute to the transmission near the three energies. When the strain is increased to $\zeta=3\%$ and $6\%$, the peak value of the transmission at $E=0.9$ eV is significantly decreased to about $1$. In particular, the transmission decreases to about $0.1$ within the energy interval of $0<E<0.15$ eV at $\zeta=6\%$, which is associated with the breaking of the energy band crossing along the S-G path shown in Fig.~\ref{fig:0Band}(c). When the strain increases further to 9$\%$, the Weyl point is completely broken. Correspondingly, the transmission peak in the energy interval $0<E<0.25$ eV drops to a small value because of the strain. In addition, the sharp transmission peak near $E=0.9$ eV evolves to a broad transmission peak, which is attributed to the accumulation of quantum states within the energy interval of $0.5<E<1$ eV observed in Fig.~\ref{fig:0Band}(d). The transmission curves can reflect the effect of strain on the energy bands obviously, including the variation of the Weyl points.
\begin{figure}[t]
\centering
\includegraphics[scale=0.28]{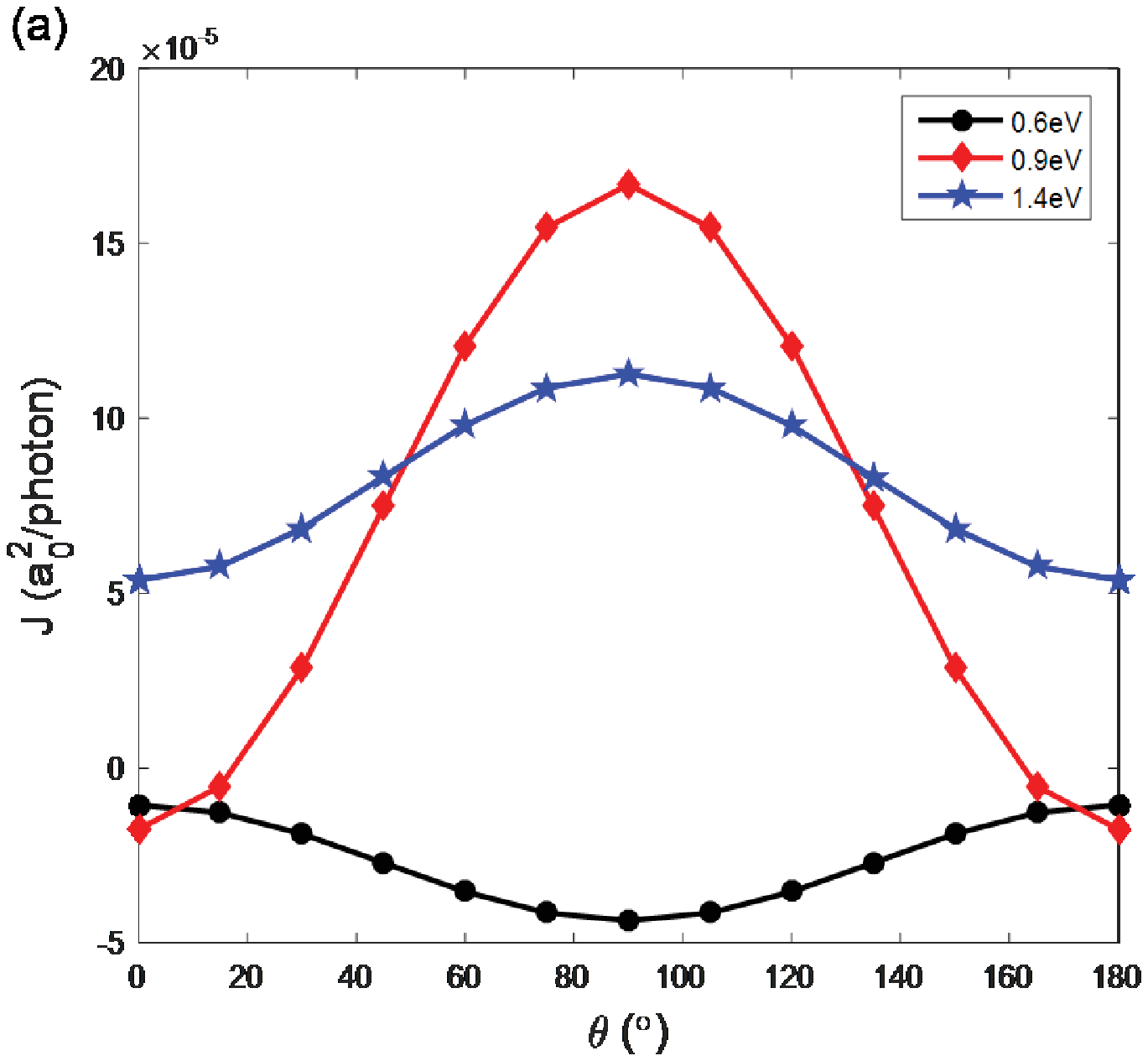}
\includegraphics[scale=0.28]{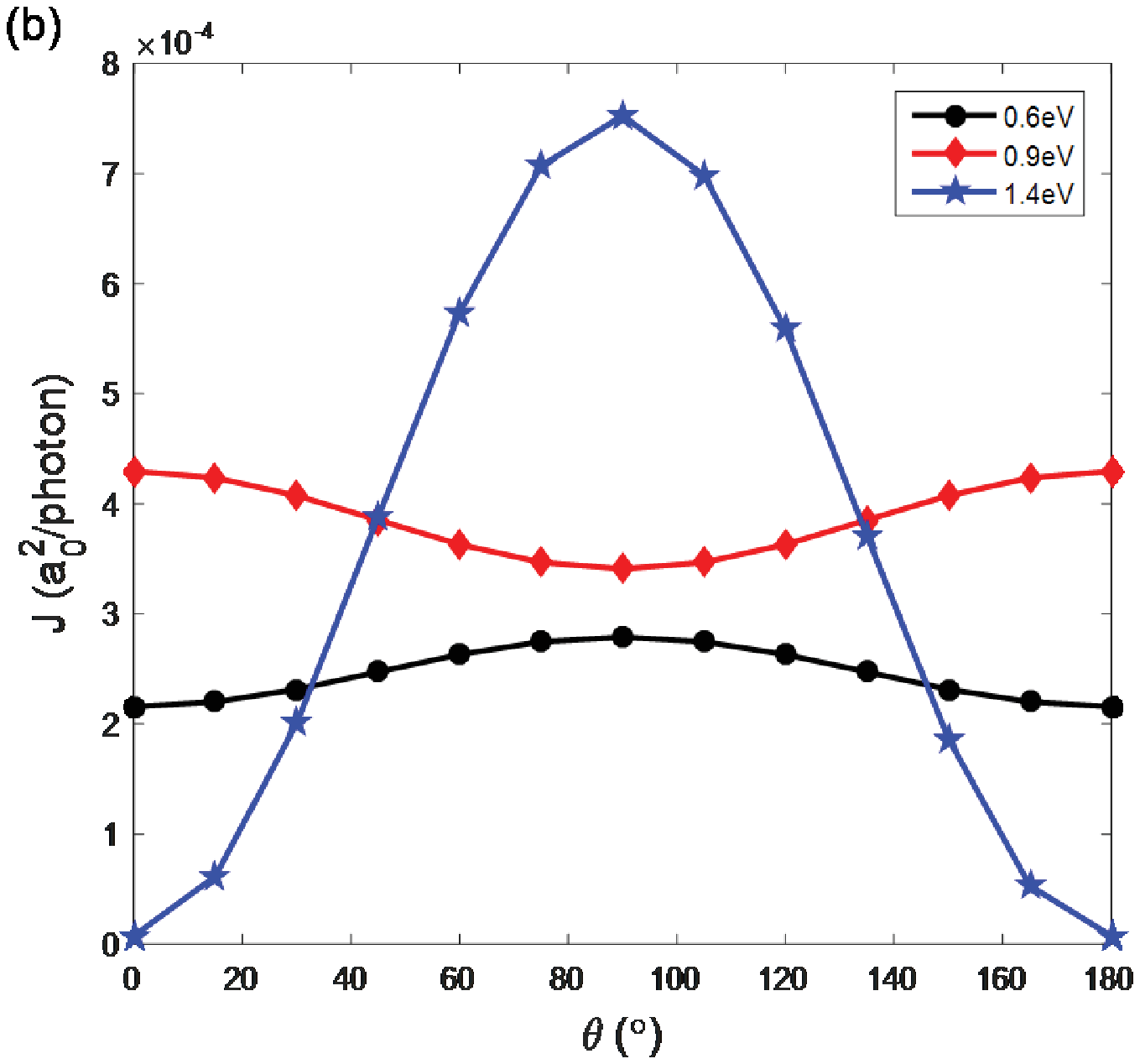}
\includegraphics[scale=0.28]{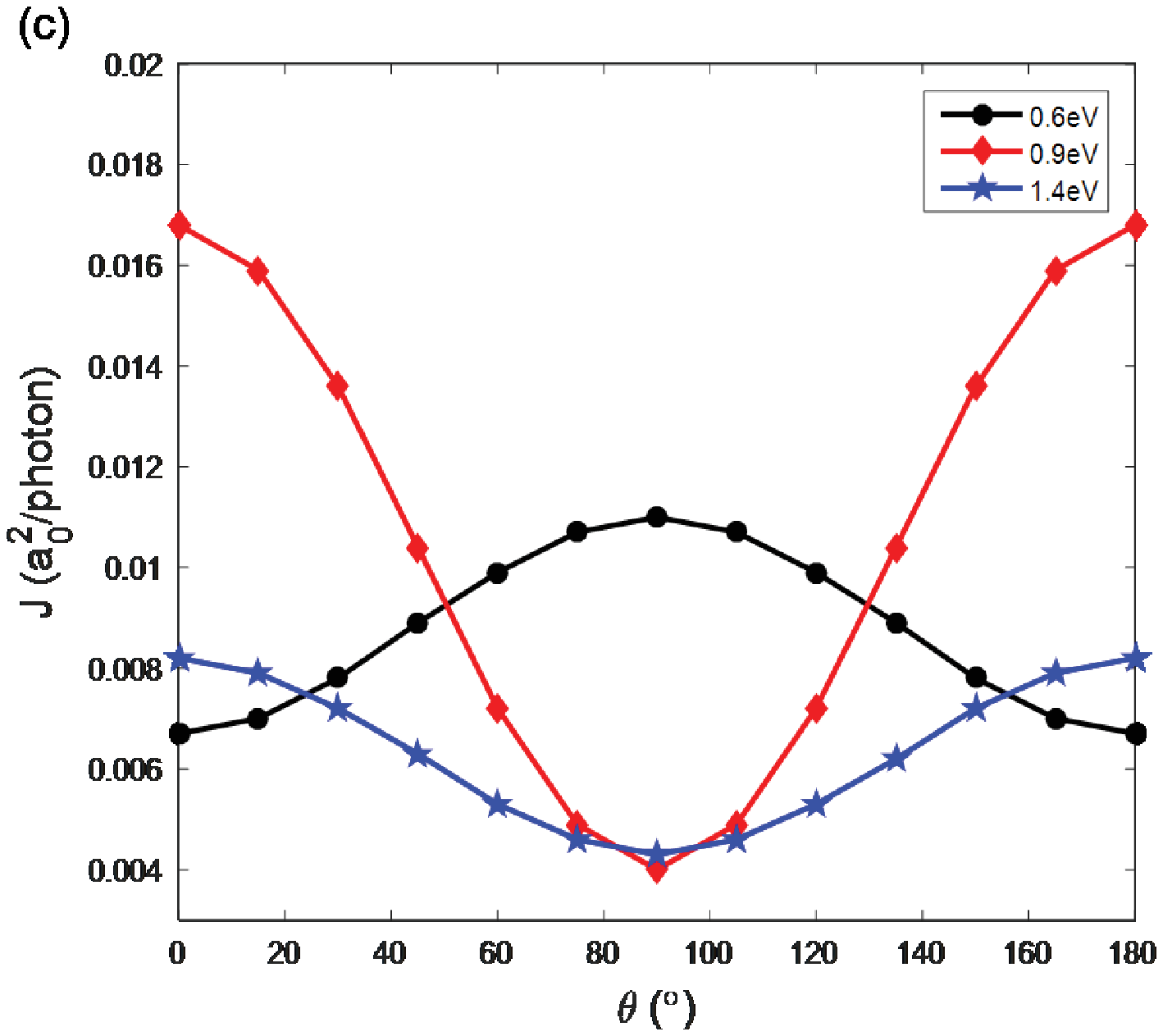}
\includegraphics[scale=0.28]{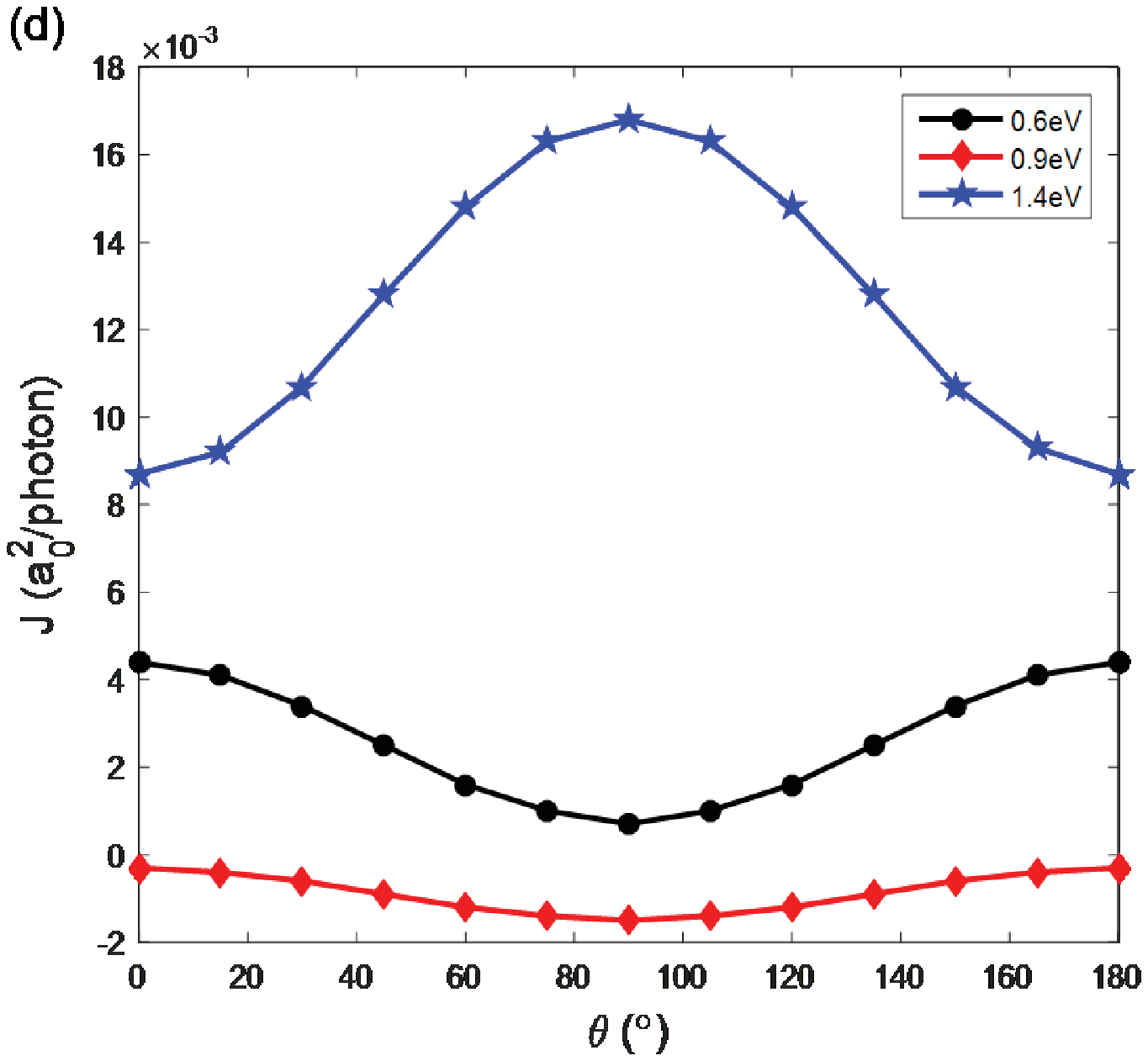}
\caption{\label{phc} The photocurrent generated in the Cu-MoTe$_2$-Cu is plotted as a function of the polarization angle of the linearly polarized light $\theta$ for photon energies from 0.6 eV to 1.4eV. The strain is applied along the $x$ direction with the amplitudes of (a) $\zeta=0$, (b) $\zeta=3\%$, (c) $\zeta=6\%$ and (d) $\zeta=9\%$.}
\end{figure}

We next analyze the photocurrent of the Cu-MoTe$_2$-Cu under varying strains along the $y$ axis to study the effect of strain on the transport characteristics of the MoTe$_2$ systems. Because single-layer T$_d$-MoTe$_2$ has no spatial inversion symmetry, the vertical irradiation of linearly polarized light can induce a large photogalvanic effect (PGE) and thereby generate photocurrent. We calculate the photocurrent for photon energies from 0.1 eV to 2.4 eV, which covers the near-infrared and visible light range. As shown in Fig.~\ref{phc}, we find that for all the photon energies considered, there is a $\cos(2\theta)$ relationship between the polarization angle $\theta$ and the photocurrent. Our calculated results for the photocurrent under $C_s$ symmetry are therefore consistent with the phenomenological theory of PGE~\cite{Belinicher,Hu,Xie2}. When $\zeta=0$, the photocurrent has a relatively large value of about $1.7\times 10^{-4}$ at $\theta=90^{\circ}$ and the photon energy 0.9 eV. All three curves exhibit the $cos(2\theta)$ dependence. As the strain increases, the maximum value of the photocurrent increases to $0.0168$ at the strain strengths of $\zeta=6\%$ and $9\%$, as shown in Figs.~\ref{phc}(c)-(d). We therefore find that for any given photon energy, the maximum values of the photocurrent are obtained at either $\theta=0^{\circ}$ or $\theta=90^{\circ}$. When strain is applied to T$_d$-MoTe$_2$ along these two directions, the cosine-curve relationship is preserved because the strain does not change the symmetry of T$_d$-MoTe$_2$. For further study, we consider the maximum value of the photocurrent $J_{max}$ at different photon energies to analyze the effect of the strain on the photocurrent.
\begin{figure}[t]
\centering
\includegraphics[scale= 0.35]{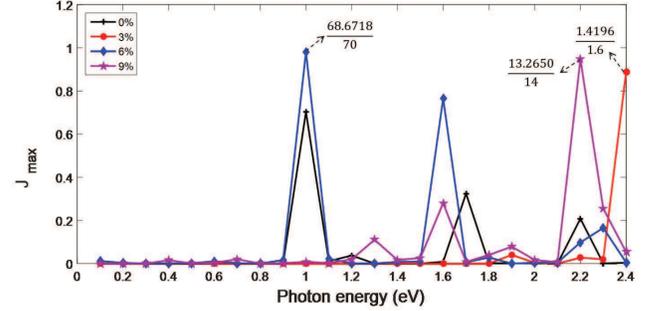}
\caption{\label{vmax} The maximum photocurrent in the Cu-MoTe$_2$-Cu induced by linearly polarized light for photon energies from 0.1 eV to 2.4 eV under strains of $\zeta=0$, $3\%$, $6\%$, and $9\%$ applied along the $y$ direction. To clearly show the variation of the photocurrent, the three peak values are scaled by a multiple. For example, $68.6718/70$ means that the peak value is
scaled to one seventieth of its actual value.}
\end{figure}

In Fig.~\ref{vmax}, the maximum photocurrent $J_{max}$ is plotted as a function of the photon energy for different strengths of the strain.
There are three peaks in $J_{max}$ in the absence of the strain. The peak values of $J_{max}$ are about $0.7$, $0.35$, and $0.2$ at the respective photon energies of $1$ eV, $1.7$ eV, and $2.2$ eV. These three peaks arise from the transition of electrons from the VB to the CB, which are
associated with the three energy spacings of $0.98$, $1.68$, and $2.28$ eV, as shown in Fig.~\ref{fig:0Band}. When $\zeta=3\%$, the largest photocurrent occurs at the photon energy of $2.4$ eV and  increases to $1.4196$. Interestingly, when the strain is increased to $\zeta=6\%$, the photocurrent increases significantly to $68.6718$ at the photon energy of $1$ eV, which is about 92 times the photocurrent at $\zeta=0$. The shift of the photon energy at which the three peak photocurrent values occur to $1$ eV, $1.6$ eV and $2.3$ eV reflects the change in the three energy spacings in Fig.~\ref{fig:0Band}(c). Especially,
when the energy spacing is $1$ eV, the energy extremas of two CBs close to each other and more quantum states can contribute to the large photocurrent.
When $\zeta=9\%$, the photocurrent has four peaks, among which the largest value of $13.265$ occurs at $2.2$ eV. The above results show that strain can be utilized to significantly enhance the photocurrent, and that the photocurrent can reflect the detailed changes in the band structure, including the modification of the Weyl point induced by the strain.
\begin{figure}[t]
\centering
\includegraphics[scale= 0.35]{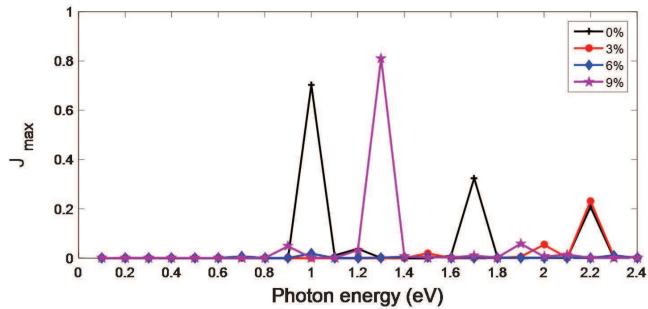}
\caption{\label{tmax} The maximum photocurrent in the Cu-MoTe$_2$-Cu induced by linearly polarized light for photon energies from 0.1 eV to 2.4 eV under strains of $\zeta=$0, 3$\%$, 6$\%$, and 9$\%$ applied along the $z$ direction. }
\end{figure}

We further study the anisotropy of the strain-induced photocurrent by considering the application of strain along the $z$ direction.
Fig.~\ref{tmax} shows the corresponding $J_{max}$ induced by linearly polarized light with photon energies from 0.1 eV to 2.4 eV when the strain is applied along the $z$ direction. When $\zeta=3\%$, there exist two photocurrent peaks at the photon energies of 2 eV and 2.2 eV. The peak value of $J_{max}$ at 2 eV is about 0.05. With the increase of the strain to $\zeta=6\%$, the photocurrent significantly decreases to zero.  When $\zeta=9\%$, two peaks occur at 1.3 eV and 1.9 eV. The peak value of $J_{max}$ at 1.3 eV increases to about 0.82, which is close to the value of the photocurrent when $\zeta=0$. The peak values of the photocurrent tends to be lower than that observed under strain which is applied in the $y$ or transport direction. The above results show that the strain-induced photocurrent exhibits a large anisotropy to the strain direction: The photocurrent can be suppressed to almost zero when the strain is exerted along the $z$ direction. In contrast, the photocurrent can be significantly increased to 92 times that of the photocurrent at $\zeta=0$ when the strain is exerted along the $y$ or transport direction. This suggests that an effective modulation of the photocurrent can thus be realized by utilizing strain.

\section{Conclusions}
In conclusion, we have studied the effect of strain on the energy band structure of T$_d$-MoTe$_2$ and the transport properties and photocurrent in a MoTe$_2$-based device.  Density functional theory was used to calculate the energy band structure and electron transmission spectrum of T$_d$-MoTe$_2$ under different strains. The results show that strain can effectively modulate the energy bands and result in the destruction of Weyl points.The transmission peaks reflect the variation of the energy bands induced by the strain.
The strain-induced photocurrent shows an anisotropic dependence on the direction of the strain which reflects the variation of the energy bands, including the Weyl point, modulated by the strain. The photocurrent can be suppressed to almost zero when the strain is exerted along the $z$ direction. In contrast, the photocurrent can be significantly increased to 92 times that of the photocurrent at $\zeta=0$ when the strain is applied along the $y$ direction. An effective modulation of the photocurrent can thus be realized by utilizing strain. These findings on the strain-modulated transmission and photocurrent are valuable for the exploration of novel MoTe$_2$-based devices, especially optoelectronic devices.

This work was supported by National Natural Science Foundation of China (Grant No. 11574067).

\textbf{Conflict of interest}
The authors have no conflicts to disclose.

\textbf{DATA AVAILABILITY}
The data that support the findings of this study are available from the corresponding author upon reasonable request.

\end{document}